\documentclass[preprint, preprintnumbers,amsmath,amssymb]{revtex4}

\usepackage{graphicx}
\usepackage{dcolumn}
\usepackage{bm}

\begin{document}

\title{Activating Mg acceptors in AlN by oxygen: first principles calculations}

\author{R. Q. Wu, L. Shen, M. Yang, Z. D. Sha and Y. P. Feng}
\email{phyfyp@nus.edu.sg}
 \affiliation{Department of Physics,
Faculty of Science, National University of Singapore, Singapore
117542}
\author{Z. G. Huang and Q. Y. Wu}
\affiliation{Department of Physics, Fujian Normal University, Fuzhou
350007, China}
\begin{abstract}
First principles calculations based on density functional theory (DFT) are performed to study the electronic properties of Mg acceptors in AlN at the presence of oxygen. It is found that Mg and O tend to form complexes like Mg-O, Mg$_2$-O, Mg$_3$-O and Mg$_4$-O which have activation energies about 0.23 eV lower than that of Mg (except of the passive Mg-O). The lower activation energies originate from the extra states over valence band top of AlN induced by the passive Mg-O. By comparing to the well-established case of GaN, it is possible to fabricate Mg and O codoped AlN without MgO precipitate. These results suggest the possibility of achieving higher hole concentration in AlN by Mg and O codoping.

\end{abstract}

\maketitle

Ultraviolet light-emitting diodes (LEDs) with wavelengths less than
300 nm are of considerable technological significance. They are
potential alternatives to the existing gas lasers and mercury lamps
in applications such as disinfection, air and water purification
and biomedicine  where the latter encounter difficulties due to their
high operating voltages, low efficiency, large size and toxicities.\cite{science}
They also promise high density optical data storage and high-resolution
photolithography.\cite{nature1} Wide gap semiconductors, diamond
($E_g$=5.5 eV) and AlN ($E_g$=6.2 eV) are the two most studied
materials for such LEDs and AlN is particularly favored due to its
direct gap band structure and subsequent high light-emission
efficiency. Diamond as well as AlN based LEDs have been fabricated
recently.\cite{science2,nature2} However, in view of their low
efficiency and high operating voltages, further developments are
still required to improve these two LEDs to the point where they can
be used as devices.\cite{nature3}

Fabrication of a homo-structured LED requires both $p$- and $n$-type
doping of a semiconductor. Unfortunately, the asymmetry dopability
of wide gap semiconductors makes the fabrication of LEDs very
difficult. The asymmetry dopability means that a wide gap
semiconductor can be either $p$- or $n$-type doped,but not
both.\cite{asymmetry} AlN can be easily $n$-type doped with Si.
However, $p$-type AlN is of great challenge. The most promising
acceptor for AlN is Mg. Yet, the activation energy ($E_A$) of the Mg
acceptor in AlN is 0.5 eV.\cite{ea1,ea2} As the ratio of
carrier concentration to impurity concentration follows
exp$(-E_A/{k_BT})$, where $k_B$ is the Boltzmann constant and $T$ is
the temperature, only a very small fraction $\sim$ 10$^{-8}$ of Mg
impurities is activated at room temperature. Since the upper limit
of Mg concentration in AlN is
2$\times$$\sim$10$^{20}$cm$^{-3}$,\cite{nature2}, the corresponding
  hole concentration for Mg-doped AlN would not exceed
$\sim$10$^{12}$cm$^{-3}$. This concentration, however, is still well
below that for device applications (which requires a hole concentration at least 10$^{17}$cm$^{-3}$  ).

Recently, codoping has been proposed and applied to overcome difficulties in
$p$-type doping in some wide gap semiconductors  such as GaN
and ZnO.\cite{jpcm} Significant improvements in hole concentrations
have been achieved in these two semiconductors. In this approach,
$p$-type dopants ($D$) are incorporated into the semiconductor along with
a small amount of reactive $n$-type impurities as codopants
($CD$). Then in the host semiconductors, complexes like $D$-$CD$,$D_2$-$CD$, $D_3$-$CD$ and $D_4$-$CD$
will form and more often they have lower ionization energies
than that of mono-dopant $D$(expect for $D$-$CD$ which is passive). Thus the hole concentration may be
greatly enhanced. By this approach, the hole concentration in ZnO was improved from $\sim$10$^{13}$cm$^{-3}$ in N-mono-doped ZnO to $\sim$10$^{17}$cm$^{-3}$ or ever higher in (Al,Ga or In)-N codoped ZnO.\cite{AlNZnO, InNZnO, GaNZnO} Codoping also enhanced the hole
concentration in Mg doped GaN. The conductivity of Mg doped GaN
could be significantly enhanced by annealing in an environment with
oxygen.\cite{GaN1} This enhancement in conductivity was attributed to
the decreased activation of Mg acceptors due to the inclusion of
oxygen and a corresponding   order of magnitude increase of hole
concentration(from $\sim$10$^{17}$cm$^{-3}$ to
$\sim$10$^{18}$cm$^{-3}$).\cite{GaN2}

It is therefore reasonable to believe that the hole concentration of Mg-doped AlN can be
improved using a similar approach.  In this letter, we
investigate the effect of atomic oxygen on the activation energy of Mg
acceptors in AlN by first principles electronic calculations  based
on density functional theory (DFT). We first study the possibility
of the formation of Mg$_n$-O  (here $n$ ranges from 1 to 4; atomic O
occupies N site and atomic Mg  the neighboring Al site,
respectively) complexes in AlN.  All these complexes under study
are assumed to be electrically neutral unless otherwise stated. This
is reasonable since even at an activation energy as low as 0.20 eV,
only a small portion ($\sim$10$^{-4}$) of the acceptors are in
charged state. Then the activation energies of {\em
complex acceptor} Mg$_n$-O is calculated. To explain the results obtained,
the density of states (DOS)  are given
 . Most of the calculations are done using a
3$\times$3$\times$2 supercell constructed from AlN
unit cell using the plane-wave DFT code VASP.\cite{vasp1,vasp2} The
lattice constants of the supercell are kept fixed to avoid effects
from spurious volume expansion. The $\Gamma$-centered
4$\times$4$\times$4 $k$-mesh is used for irreducible Brillouin zone
sampling. The ion-electron interaction is described by Vanderbilt
ultrasoft pseudopotentials\cite{uspp}with local density
approximation (LDA) for the exchange-correlation potential. The
electron wave function is expanded in plane waves with a cutoff
energy of 400 eV. These parameters ensure a convergence better than
1 meV for the total energy. In all the doped supercells, atomic
coordinates are fully relaxed using the conjugate-gradient
algorithm\cite{conjugate} until the maximum force on a single atom
is less than 0.03 eV/\AA.


For AlN in wurtzite structure the calculated lattice constants are
3.08 \AA\ for $a$ and 4.94 \AA\ for $c$ with the internal parameter
being 0.382, in good agreement with experimental values. Based on this the
3$\times$3$\times$2 supercell (Al$_{36}$N$_{36 }$) is constructed. Previous calculations on formation energies have shown that O occupies N site in AlN as a deep donor\cite{on1,on2} and consequently in our supercell one N atom is replaced by an O atom. This O
donor may act as an attraction center to single Mg atom and form
complexes such as Mg-O, Mg$_2$-O, Mg$_3$-O and Mg$_4$-O with Mg
atoms occupying the nearest Al sites to O and without destroying the
lattice structure. To study whether single Mg atom will bind to
Mg$_n$-O ($n$=0,1, 2 and 3, respectively) complexes, we define the
{\em binding energy} as the energy required to form the Mg$_{n+1}$-O complex from well separated Mg dopant and   Mg$_n$-O complex:
\begin{eqnarray}
\Delta^{(n)}=E(Al_{36-n-1}Mg_{n+1}N_{35}O)+E(Al_{36}N_{36})\nonumber\\
-E(Al_{35}Mg_{1}N_{36})-E(Al_{36-n}Mg_{n}N_{35}O)
\end{eqnarray}
where $E$ is the total energy of the system indicated in
parentheses. A negative $\Delta^{(n+1)}$ suggests that the
Mg$_{n+1}$-O complex is energetically favorable and stable while a
positive $\Delta^{(n)}$ suggests that Mg$_{n+1}$-O cannot form.  The
calculated $\Delta^{(n)}$ are summarized in Table ~\ref{tab1}. As
can be seen, single Mg atom will bind to single O atom for a
large energy decrease of 5.304 eV. This large energy decrease can be
attributed to the passivation of the extra electron of O by Mg. The
energy decrease from single Mg atom binding to Mg-O complex is 0.626
eV, suggesting that Mg$_2$-O complex will form provided the amount
of Mg exceeds that of O. If there are still extra Mg acceptors
available then Mg$_3$-O will form with an energy decrease of 0.412 eV
. Formation of Mg$_4$-O
from single Mg atom and Mg$_3$-O is less likely although also
exothermic due to a much smaller energy decrease of 0.157 eV. The
binding of single Mg atom to the O atom in Mg$_n$-O complexes can be
attributed to the larger electronegativity of O than that of N.

Now we calculate  the activation energy of the Mg$_2$-O complex
which is the most probable complex when the concentration ration between Mg and O is
 around 2. The method proposed by Van de Walle is
applied for the activation calculation:\cite{ea}
\begin{eqnarray}
E{_A}=E{_{tot}}[D^{-}]-E{_{tot}}[D^{0}]-E_{v}-\Delta V[D]+E_{corr}
\end{eqnarray}
where $E_A$ is the activation energy of the defect (donor) $D$.
$E{_{tot}}[D^{-}]$ and $E{_{tot}}[D^{0}]$ are the total energies of
the supercell with defect in charged (-) and neutral (0) state,
respectively. $E_v$ is the valence band maximum of the bulk
semiconductor; $\Delta V[D]$ is a correction term to align the
reference potential in the charged defected supercell with that of
the bulk and it is derived from the electrostatic potential
difference between the bulk and  that of the defected supercell far
away from defect site. Using a 5$\times$5$\times$3 supercell a value
of -0.15 eV is calculated for both $\Delta V[Mg]$ and $\Delta
V[Mg_2-O]$. Calculation obtained using  a larger 6$\times$6$\times$4 supercell
gives out does not result in any significant change in these values. $E_{corr}$ is a correction term
for the use of $\Gamma$-included $k$-mesh sampling for the hexagonal
lattice. In practice this is derived from the energy difference
between the highest occupied level at $\Gamma$-point and other
special $k$-points (averaged) in the supercell containing the
neutral defect $D$. The calculated $E_{corr}$ is 0.26 eV for
supercell containing only Mg defect and 0.03 eV for the supercell
containing Mg$_2$-O complex. The calculated ionization of single Mg
acceptor in AlN is 0.40 eV which is in agreement with previous
DFT-LDA calculation ($\sim$ 0.45 eV \cite{jap}) and experiments
($\sim$ 0.5 eV\cite{ea1,ea2}). For the Mg$_2$-O complex, the
calculated $E_a$ is only 0.17 eV, 0.23 eV lower than that of single
Mg acceptor. If we apply this decrease to the experimental value,
then following exp$(-E_A/{k_BT})$, the hole concentration can be
increased by at least a factor of 10$^{3}$. This is a significant
improvement, although the total carrier concentration is still below
the value desired.  Other clusters Mg$_3$-O and Mg$_4$-O have very close activation energies to that of Mg$_2$-O. Thus incorporation of some amount of oxygen into Mg-doped AlN can improve the hole concentration.  

The decreased activation energy in the Mg acceptor after attaching
to O-Mg complex  can be understood from the density of states of the
defected supercell as shown in Figure.~\ref{fig2}. Although O-Mg
complex is passive and cannot accept the host valence electrons, it
induces extra fully occupied states right on the valence band
maximum (VBM) as indicated by the DOS curve of AlN supercell
containing Mg-O complex. In the Mg-O complex some electrons have
higher energy than that of the host valence electrons. Thus Mg atoms
binding to this complex may be activated by electrons from these
{\em complex states} rather than from the host states. So the
activation energy is decreased.

However, for such a codoping the growth condition should be chosen carefully. This is because under thermal equilibrium the formation energies of Mg$_n$O complexes will be rather high. The formation energy is given by:
\begin{eqnarray}
E^f[D(q)]=E^{tot}[D(q)]-E^{tot}[bulk]-\Sigma_{i} {n_{i}\mu_i}\nonumber \\
+q(E_f+E_v)
\end{eqnarray}
where the $E_{defect}^{tot}(q)$ is the total energy of the supercell containing the defect,
$E^{tot}(bulk)$ is the total energy of a similar supercell containing the pure crystal, $n{_i}$ is the number of atoms that is involved in the formation of the defect with $\mu_i$ being the corresponding chemical potentials. $E_f$ is the Fermi energy which is set to zero at the valence-band maximum $E_v$. The chemical potentials depend on the experimental growth conditions, which can be either Al-rich or N-rich. Formation of AlN crystal under thermal equilibrium requires
$\mu_{Al}$+$\mu_N$=$\mu_{Al}[bulk]$+$\frac1 2 \mu_N[N_2]$+$\Delta H[AlN]$, where $\Delta H(AlN)$ is the formation enthalpy of AlN. In the N-rich condition which is preferred for incorporating Mg at Al sites, the upper limit of $\mu_N$ is given by $\mu_N[N_2]$, i.e., the energy of N in a N$_2$ molecular. $\Delta H[AlN]$ is calculated to be -3.58 eV.  The formation energy of Mg on Al site at neutral state is $E^f[Mg_{Al}(0)]$=1.76 eV with the solubility limit imposed by 3$\mu_{Mg}$+2$\mu_N$=3$\mu_{Mg}[bulk]$+2$\mu_N[N_2]$+ $\Delta H(Mg_3N_2)$. However, if O$_2$ is present in N$_2$ flux, to avoid the formation of MgO precipitate the upper limit of $\mu_{Mg}$ follows:
\begin{eqnarray}
\mu_{Mg}+\mu_O=\mu_{Mg}[bulk]+\mu_O[O_2]+\Delta H(MgO) \nonumber
\end{eqnarray}
where $\Delta H(MgO)$=-6.69 eV from our calculation. Under this constrain the formation energies of Mg$n$-O complexes become rather high. The formation energy of Mg$_2$-O complex is calculated to be 4.81 eV, suggesting an ignorable concentration of Mg$_2$-O complex in AlN in a growth process which is close to thermal equilibrium. However, this solubility limit can be overcome by proper growth conditions, i.e. high growth temperature and high growth rate. For example, in the doping of GaN with Mg as acceptors by gas-source epitaxy method, the inclusion of oxygen did not result in the MgO precipitate in GaN while improved the hole concentration significantly.\cite{GaN2,mgo2} By calculation, the formation energy of Mg$_2$-O complex in GaN is 6.31 eV, significantly higher than that of AlN. Thus in AlN, the solubility limit problem is less severe. This suggests the possibility of Mg and O codoping in AlN without MgO precipitates.

To summarize, we have studied the electronic properties of Mg acceptors
in AlN at the presence of oxygen by first principles calculations. Our calculations suggest the formation of of Mg$_n$-O complexes and their lower activation energies compared to Mg. Compared to the well established case of GaN, the MgO precipitate problem can be overcome. Our results suggest that the hole concentration in AlN:Mg can be greatly enhanced by oxygen codoping.

This work was partly supported by NSF of China under Grant
No.6067655 and National Key Project for Basic Research of China
under Grant No. 2005CB623605.

\clearpage
\begin{table}
\caption {{\em Binding energy} for single Mg acceptor to Mg$_n$-O complex
 . Unit: eV} \centering
\begin{tabular}{ l c c c c}
\hline \rule{0mm}{0mm}$n$\rule{6mm}{0mm} & \rule{6mm}{0mm}0
 \rule{6mm}{0mm} &
\rule{6mm}{0mm}1\rule{6mm}{0mm} &\rule{6mm} {0mm} 2 \rule{6mm} {0mm} &\rule{6mm} {0mm} 3 \rule{6mm} {0mm}\\
\hline
$\Delta^{(n)}$ & -5.304 & -0.626 &  -0.412 & -0.157\\
\hline
\end{tabular}
\label{tab1}
\end{table}
\clearpage

 \begin{figure}
 \caption{Total DOS of supercells with pure AlN [curve marked by AlN]
and AlN containing one passive Mg-O complex [curve marked by
AlN(Mg-O)]. The bold arrow indicates the extra states on the top of
the VBM of the pure AlN. The Fermi level of the pure AlN is
indicated by the vertical dashed line.} \label{fig2}
\end{figure}

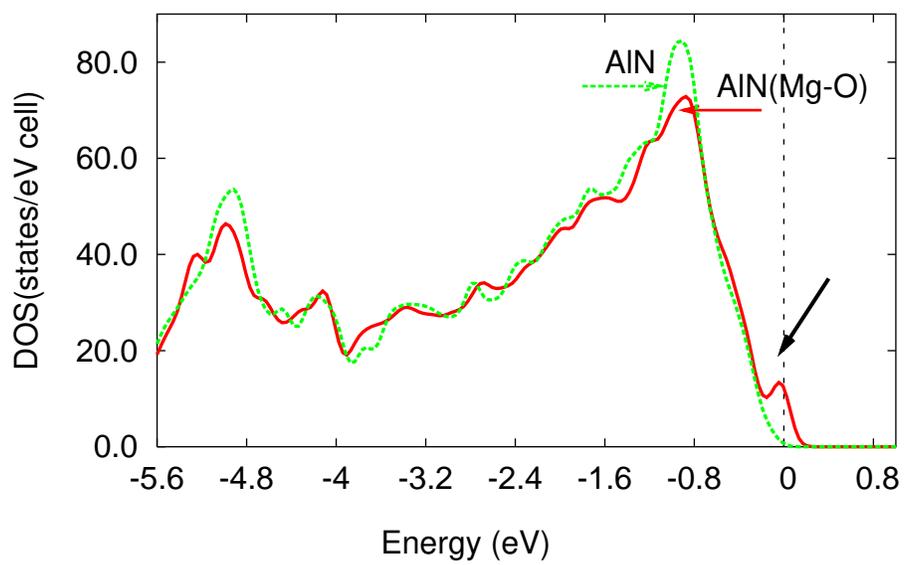
\end{document}